\documentclass[12pt]{article}
\usepackage{cite,graphicx}

\newcommand{\newc}{\newcommand}
\newc{\gsim}{\lower.7ex\hbox{$\;\stackrel{\textstyle>}{\sim}\;$}}
\newc{\lsim}{\lower.7ex\hbox{$\;\stackrel{\textstyle<}{\sim}\;$}}
\newc{\gev}{\,{\rm GeV}}
\newc{\mev}{\,{\rm MeV}}
\newc{\ev}{\,{\rm eV}}
\newc{\kev}{\,{\rm keV}}
\newc{\tev}{\,{\rm TeV}}

\newc{\mz}{M_Z}
\newc{\mpl}{M_*}
\newc{\mw}{m_{\rm weak}}
\def\beq{\begin{equation}}
\def\eeq{\end{equation}}
\def\bea{\begin{eqnarray}}
\def\eea{\end{eqnarray}}
\newc{\ie}{{\it i.e.}}          \newc{\etal}{{\it et al.}}
\newc{\eg}{{\it e.g.}}          \newc{\etc}{{\it etc.}}
\newc{\cf}{{\it c.f.}}
\def\bar#1{\overline{#1}}
\def\vev#1{\left\langle #1 \right\rangle}

\def\inv{^{\raise.15ex\hbox{${\scriptscriptstyle -}$}\kern-.05em 1}}
\def\lbar{{\lower.35ex\hbox{$\mathchar'26$}\mkern-10mu\lambda}} %lambda bar

\let\<=\langle
\let\>=\rangle

\let\+=\uparrow
\let\al=\alpha
\let\be=\beta

\let\de=\delta

\let\ep=\varepsilon

\let\la=\lambda

\let\si=\sigma
\let\Si=\Sigma
\let\th=\theta

\renewcommand{\epsilon}{\varepsilon}
\renewcommand{\phi}{\varphi}

\def\t#1{T_{#1}}

\def\fb#1{{\bar F}_{#1}}

\begin{document}
\thispagestyle{empty}
\vspace*{.5cm}
\noindent
\hspace*{\fill}{\large CERN-TH/2001-194}\\
\hspace*{\fill}{\large UCB-PTH-01/34}\\
\hspace*{\fill}{\large LBNL-48632}
\vspace*{2.0cm}

\begin{center}
{\Large\bf Towards a Theory of Flavor from Orbifold GUTs}
\\[2.5cm]
{\large Lawrence Hall$^{1,2}$, John March-Russell$^{1,3}$,
Takemichi Okui$^{1,2}$,}\\[.2cm]
{\large and~David~Smith$^1$
}\\[.5cm]
{\it $^1$Theory Group, Physics Division, LBNL, Berkeley, CA 94720, USA}
\\[.2cm]
{\it $^2$Physics Department, University of California, Berkeley,
CA 94720, USA}
\\[.2cm]
{\it $^3$Theory Division, CERN, CH-1211 Geneva 23, Switzerland}
\\[.2cm]
(August, 2001)
\\[1.1cm]

{\bf Abstract}\end{center}
\noindent
We show that the recently constructed 5-dimensional supersymmetric
$S^1/(Z_2\times Z_2')$ orbifold GUT models allow an appealing explanation
of the observed hierarchical structure of the quark and lepton masses
and mixing angles. Flavor hierarchies arise from the geometrical
suppression of some couplings when fields propagate in different numbers
of dimensions, or on different fixed branes. Restrictions arising from
locality in the extra dimension allow interesting texture zeroes to
be easily generated. In addition the detailed nature of the SU(5)-breaking
orbifold projections lead to simple theories where $b-\tau$ unification
is maintained but similar disfavored SU(5) relations for the lighter
generations are naturally avoided.  We find that simple 5d models based
on $S^1/(Z_2\times Z_2')$ are strikingly successful in explaining many
features of the masses and mixing angles of the 2nd and 3rd generation. 
Successful three generation models of flavor including neutrinos
are constructed by generalizing the $S^1/(Z_2\times Z'_2)$ model to
six dimensions. Large angle neutrino mixing is elegantly accommodated.
Novel features of these models include a simple $m_u=0$ configuration
leading to a solution of the strong CP problem. 
\newpage

\setcounter{page}{1}
\section{Introduction}

Since the pioneering work of Georgi and Glashow~\cite{GG} (also
\cite{PS}) the compelling concept of a grand unified theory (GUT)
of the Standard Model (SM) gauge interactions has dominated
our thinking about physics at high energies.  The quantitative
success of gauge coupling unification~\cite{GQW} in the minimal
supersymmetric standard model (MSSM) extensions of these
theories~\cite{DRW,DG} has provided further support 
for this idea~\cite{unif}.  

These successes motivate one to look for similar high-scale
explanations of the rich flavor structure we observe at low energies. 
While the three Yukawa matrices appear as independent parameters
in the Standard Model, they can become correlated in an extension:
\eg, in specific GUTs there are relations between fermion masses,
such as the SU(5) relation $m_b=m_\tau$~\cite{BEGN}.  As with the
unification of gauge couplings this
relation applies only at the unification scale and is modified by
renormalization group (RG) running down to the weak scale. Indeed, these
corrections offer a further tantalising piece of circumstantial
evidence in favour of supersymmetric unification for, starting with
the relation $m_b(M_X)=m_{\tau}(M_X)$, they bring the prediction for
$m_b/m_\tau$ into agreement with experiment using the same value
for the unification scale $M_X$ as found in the analysis of gauge
couplings.  This illustrates both how the analysis of fermion masses
can lend support to the hypothesis of (supersymmetric) unification,
and how we can learn about the puzzling flavor structure of the SM
from unified theories.

However, additional predictions for fermion masses and mixing angles
require more sophisticated unified theories.  One reason for this
is that the simple relations $m_s=m_\mu$ and $m_d=m_e$
are in gross disagreement with experiment even with RG corrections
included.  Such problems inspired Georgi and Jarlskog~\cite{GJ}
to study a different unified ansatz, which was then developed
and extended to other ans\"atze in many subsequent studies with some
success (see for example~\cite{FN,textures}).  Despite
this, one problem that looms over GUTs, and especially their extensions
addressing such flavor structure, is the difficulty of building a Higgs
structure that achieves the desired pattern of vacuum expectation
values (VEVs) and breaking.  Most notorious is the doublet-triplet
splitting problem of SU(5) and other GUTs.  Although solutions, such
as the Dimopoulos-Wilczek mechanism in SO(10) are possible~\cite{BB},
the full Higgs structure necessary to realise the flavor relations
tends to be unattractive.  In addition the constraints from the
non-observation of proton decay are now so severe as to disfavor
or even rule out many of the simplest models~\cite{MP}.

Recently, a new possibility for the embedding of the SM into a form of
GUT has been suggested in
Refs.~\cite{kaw,AF,HN,HMR,HMR2,recent,HNOS}.  The idea is 
that the GUT gauge symmetry is realized in 5 or more space-time
dimensions and only broken down to the SM by compactification on a 
singular `orbifold', utilizing GUT-symmetry violating boundary conditions. 
Given the success of the traditional supersymmetric gauge-coupling
unification predictions, the most attractive model is one with both
supersymmetry and (in a sense we make precise) SU(5) gauge symmetry.
Then in the 5d case both the
GUT group and 5d supersymmetry are broken by compactification on
$S^1/(Z_2\times Z_2')$ down to a N=1 SUSY model with SM gauge group.
This construction allows one to avoid some unsatisfactory features
of conventional GUT's with Higgs breaking, such as doublet-triplet
splitting, while maintaining, at least at leading order, the desired
gauge coupling unification~\cite{HN,HMR,CPRT}\footnote{These features
are shared with those of the original string orbifold
constructions~\cite{string}.}.  The significant advantage of the effective
field theory approach to orbifolds that we advocate, is that large classes
of (consistent low-energy) models can be surveyed, and phenomenologically
interesting features or mechanisms identified. 

In this paper we examine what we consider a very appealing explanation
of the hierarchical structure of the quark and lepton masses and
mixing angles based on a generalization of the $S^1/(Z_2\times Z_2')$
orbifold GUT model.  Our hypothesis is that flavor
hierarchies are a result of the geometrical suppression of some
couplings due to wavefunction normalization factors arising
when fields propagate in different numbers of dimensions. 
Locality and the detailed nature of the orbifold projections
in field space play significant roles as fixed branes
of various dimensionalities and in various locations are an
automatic feature of the $S^1/(Z_2\times Z_2')$ orbifold mechanism
for SU(5) breaking.  We find that quite simple models of this sort
lead to notably successful flavor textures.

The outline of our paper is as follows.
In Section~2 we briefly recall the essential features of the
$S^1/(Z_2\times Z_2')$ orbifold mechanism, paying particular
attention to aspects that are important in the construction
of our theory of flavor.  In Section~3 we explain the basic idea
of our approach to the flavor problem, while Section~4 is devoted
to the examination of these ideas in the simplified
context of models for the
flavor structure of the heaviest 2 generations.  We find that
a simple hypothesis is strikingly successful in explaining many
features of 2nd and 3rd generation masses and mixing angles:
the only matter propagating in the 5th dimension is the second
generation ${\bf 10}$ of SU(5).  Full three generation models
of flavor require that we generalize this $S^1/(Z_2\times Z'_2)$
model to six dimensions.  The construction and appealing aspects
of such models are discussed in Section~5. 
Finally, Section~6 contains our conclusions.

\section{Basics of the $S^1/Z_2\times Z'_2$ model}

In this Section we provide an introduction to the physics
of the simplest orbifold GUT model, the SU(5) model on the
5 dimensional space $M_4\times(S^1/Z_2\times Z'_2)$.  We closely
follow the discussion and notation of Refs.~\cite{HN,HMR}.

In addition to 4d Minkowski space consider a 5th dimension with
orbifold structure $S^1/(Z_2\times Z_2')$.  (We label
the usual 4d coordinates $x^{\mu}$, $\mu=0,\ldots,3$, while the 5th
coordinate $x^5\equiv y$.) The circle $S^1$ is assumed to have radius
$R$ where $1/R\sim M_{\rm GUT}$.
The orbifold $S^1/(Z_2\times Z_2')$ is obtained by modding out the theory
by two $Z_2$ transformations which impose on bulk fields the equivalence
relations: $P:~~y\sim -y$ and $P':~~y' \sim -y'$ 
(here $y'\equiv y+\pi R/2$).  Under these actions
there are two inequivalent fixed 3-branes (or `orbifold planes'),
which we denote $O$ and $O'$, located at $y=0$, and $y =\pi R/2\equiv\ell$, 
respectively.  The physical domain of the theory is the interval
$y\in [0,\ell]$ with the $O,O'$ branes acting as `end-of-the-world' branes.

The action of the equivalences $P,P'$ must also be defined within the
space of fields: $P: \Phi(x,y)\sim P_\Phi \Phi(x,-y)$ and
$P': \Phi(x,y')\sim  P'_\Phi \Phi(x,-y')$, where
here, $P_\Phi$ and $P'_\Phi$ are matrix representations
of the two $Z_2$ actions, and we can classify the fields by their
$(P,P')$ eigenvalues $(\pm 1,\pm 1)$.  Then the KK expansions of bulk
fields $\Phi_{PP'}(x,y)$ involve $\cos(k y/ R)$ with $k=2n$ or $2n+1$,
for $\Phi_{+P'}$ ($P'=+,-$
respectively), and $\sin(ky/R)$ with $k=2n+1$ or $2n+2$, for
$\Phi_{-P'}$ ($P'=+,-$ respectively).
Only the $\Phi_{++}$ possess a massless zero mode,
the other KK modes acquire a mass $k/R$ from the 4d perspective.
Only $\Phi_{++}$ and $\Phi_{+-}$
have non-zero values at $y=0$, while only $\Phi_{++}$ and $\Phi_{-+}$
are non-vanishing at $y=\ell$.  The action of the identifications
$P,P'$ on the fields (the matrices $P_\Phi$ and $P'_\Phi$)
can involve any symmetry of the bulk theory; gauge transformations,
discrete parity transformations, and R-symmetry transformations in the
supersymmetric case; and this allows one to break the bulk symmetries.

The starting point for phenomenology is a 5d SU(5) gauge theory with
minimal 5d SUSY (8 real supercharges, corresponding to $N=2$ SUSY in 4d).  
Thus, at minimum, the
bulk must have the 5d vector superfield, which in terms of 4d $N=1$ SUSY
language contains a vector supermultiplet $V$ with physical components
$A_\mu,\la$, and a chiral multiplet $\Si$ with components $\psi,\si$.
Both $V$ and $\Si$ transform in the adjoint representation of SU(5).
If the parity assignments, expressed in the fundamental representation
of SU(5), are chosen to be
$P={\rm diag}(+1,+1,+1,+1,+1)$, and $P'={\rm diag}(-1,-1,-1,+1,+1)$,
so that the equivalence under $P$ is 
\beq
V^a(x,y)T^a \sim V^a(x,-y)P T^a P^{-1} ,
\label{eq:PonV}
\eeq
and similarly for $P'$, then SU(5) is broken
to SU(3)$\times$SU(2)$\times$U(1) on the $O'$ fixed-brane, but is
unbroken in the bulk and on $O$.
If for $\Si$ the same assignments are taken apart from an overall sign
for both $P$ and $P'$ equivalences, \eg, under $P$,
\beq
\Si^a(x,y)T^a  \sim - \Si^a(x,-y) P T^a P^{-1} ,
\label{eq:PonSi}
\eeq
then these boundary conditions also break 4d $N=2$ SUSY to 4d $N=1$
SUSY on both the $O$ and $O'$ branes.  Only the $(+,+)$ fields
$V^a$ possess massless zero modes ($a$ labels the unbroken
SU(3)$\times$SU(2)$\times$U(1) generators of SU(5), while $\hat{a}$
labels the broken generators), and at low energies only the gauge
content of the 4d $N=1$ MSSM is apparent.

On the other hand, the bulk of the theory is invariant under both
the full SU(5) gauge symmetry and the full 5d minimal supersymmetry. 
If we take the MSSM matter to reside in the bulk, then they must come
in complete SU(5) multiplets.  In fact the correct situation with
regard to quantum numbers is slightly more subtle.  The reason for this
is that the bulk $N=2$ SUSY appears to pose a problem.  The minimal
matter superfield representation for such a theory is a hypermultiplet,
which in 4d N=1 language decomposes in to a chiral multiplet $\Phi_{\bf R}$
together with a mirror chiral multiplet in the conjugate representation
$\Phi^c_{\bf \bar{R}}$.  Thus, the choice of matter in the bulk would
appear to have problems reproducing the chiral structure of the SM.

Fortunately as shown in detail in Ref.~\cite{HMR} (see also \cite{HN})
the structure of the orbifold projections $P$ and $P'$ acting on
fields resolves this in a particularly interesting fashion.  The
action of these
projections on the N=1 component fields $\Phi$ and $\Phi^c$ residing in a
5d hypermultiplet is inherited from the action on the 5d vector
multiplet Eqs.~(\ref{eq:PonV})-(\ref{eq:PonSi}).
The result is that actions of both $P$ and $P'$ on the 4d chiral
fields $\Phi$ and $\Phi^c$ have a relative sign:
\beq
P: \Phi \sim P_\Phi\Phi , \qquad P: \Phi^c \sim -P_\Phi\Phi^c\,,
\label{eq:Ponhypers}
\eeq
and similarly for $P'$.  This difference leads to a chiral spectrum
for the zero modes.  Indeed the KK spectrum of 5d bulk hypermultiplets
in the representations ${\bf 10}$ and ${\bf \bar{5}}$ (whose 4d chiral
components we denote $T+ T^c +\bar{F}+\bar{F}^c$) resulting from
the $P,P'$ actions is given in Table~1.  Note that since $P'$
does not commute with SU(5), components in different
SU(3)$\times$SU(2)$\times$U(1) representations have
different KK mode structures.  Thus $Q,\bar{U},\bar{D},L,\bar{E}$,
etc., are used to indicate their SM transformation properties.
It is important in the following that the zero modes under this action
{\em do not fill out full} SU(5) {\em multiplets} of SM matter.  From
$T$ we just get ${\bar U}$ and ${\bar E}$ zero modes, while from
$\bar{F}$ we get $L$.

\begin{center}
\begin{tabular}{|c|c|c|}
\hline
$(P,P')$ & 4d superfield & 4d mass\\
\hline
$(+,+)$ &  $T_{\bar U}, T_{\bar E}, \bar{F}_{L}$  & $2n/R$\\
$(+,-)$ &  $T_{Q}, \bar{F}_{\bar D} $ & $(2n+1)/R$ \\
$(-,+)$ &  $T^c_{\bar Q}, \bar{F}^c_{D}$ & $(2n+1)/R$\\
$(-,-)$ &  $T^c_{U}, T^c_{E}, \bar{F}^c_{\bar L} $ & $(2n+2)/R$ \\
\hline
\end{tabular}
\end{center}
\vspace{3mm}
Table~1. Parity assignments and KK masses of
fields in the 4d chiral
supermultiplets resulting from the decomposition of 5d hypermultiplets in
the $(T+\bar{F})$ representation. The subscript labels the SM transformation
properties, \eg, $Q=({\bf 3},{\bf 2})_{\bf 1/6}$,
$\bar{Q}=(\bar{{\bf 3}},\bar{{\bf 2}})_{-{\bf 1/6}}$,
$\bar{U}=(\bar{\bf 3},{\bf 1})_{-{\bf 2/3}}$, etc.

The remaining components of a full SU(5) multiplet are realised
at the zero mode level by taking another copy of ${\bf 10}$ and/or
${\bf \bar{5}}$ in the bulk (with $N=1$ chiral components denoted $T'+T'^c$
and $\bar{F'}+\bar{F'}^c$ respectively), and using the freedom
to flip the overall action of the $P'$ parity on these multiplets by a
sign relative to the action on $T+\bar{F}$.
This difference leads to a different selection of zero mode components,
the KK spectrum being given in Table~2.

\begin{center}
\begin{tabular}{|c|c|c|}
\hline
$(P,P')$ & 4d superfield & 4d mass\\
\hline
$(+,+)$ &  $T'_{Q}, \bar{F'}_{\bar D} $ & $2n/R$ \\
$(+,-)$ &  $T'_{\bar U}, T'_{\bar E}, \bar{F'}_{L}$  & $(2n+1)/R$\\
$(-,+)$ &  $T'^c_{U}, T'^c_{E}, \bar{F'}^c_{\bar L} $ & $(2n+1)/R$ \\
$(-,-)$ &  $T'^c_{\bar Q}, \bar{F'}^c_{D}$ & $(2n+2)/R$\\
\hline
\end{tabular}
\end{center}
\vspace{3mm}
Table~2. Parity assignments and KK masses
of fields in 4d chiral
supermultiplets resulting from the decomposition of 5d hypermultiplets
in the $(T'+\bar{F'})$ representation. 

Combining the results of Tables~1 and 2
we have zero modes which fill out the full matter content of
either (or both) a ${\bf 10}$ or $\bar{\bf 5}$ multiplet of SU(5)
at the zero mode level, but without the doubling due to $N=2$ bulk
SUSY.  Moreover, since different components of what we think
of as a single ${\bf 10}$ or $\bar{\bf 5}$ in fact arise
from different parent multiplets in the higher dimension,
the usual SU(5) counting of independent couplings,
and thus the relations between masses and mixing angles
is modified.  We will soon utilize this feature in the construction
of realistic models of flavor.

\section{Flavor Hierarchies from Geometry}

First, to set notation, recall that in the Standard Model, the quark
and lepton masses and the Cabibbo-Kobayashi-Maskawa (CKM) mixing matrix
are parameterized in terms of three $3\times 3$ matrices, $(U,D,E)$ of
Yukawa coupling constants.  In the MSSM, and after electroweak symmetry
breaking the Yukawa interactions lead to:
\beq
{\cal L}_{\rm mass}={\bar Q}^i_L U_{ij} u^j_R {v \over \sqrt{2}} \sin\be
+{\bar Q}^i_L D_{ij} d^j_R {v \over \sqrt{2}} \cos\be
+{\bar L}^i_L E_{ij} e^j_R {v \over \sqrt{2}} \cos\be + {\rm h.c.},
\label{eq:yukawa}
\eeq
where $i,j=1,2,3$ are generation indices, $\tan\be=\vev{H_u}/\vev{H_d}$,
is the ratio of Higgs vacuum expectation values, and $v=246\gev$.
The full Yukawa matrices, $U$ and $D$ are not 
determined by experiments, rather they fix only the diagonal
mass matrices, $U^{\rm diag}$, and
$D^{\rm diag}$, and the CKM mixing matrix, $V_{CKM}$.
The relations between these objects are specified by the diagonalizing
unitary rotations $L_u, R_u, L_d$, and $R_d$ acting in generation space
on the up and down sectors: 
\beq
U^{\rm diag}=L_u U R_u^\dagger , \qquad
D^{\rm diag}=L_d D R_d^\dagger , \qquad
V_{CKM}=L_u L_d^\dagger .
\label{eq:relations}
\eeq

Our basic strategy to go beyond the usual SU(5) treatment of
the flavor structure of the SM matter is to use the geometrical
suppression of some couplings due to the wavefunction normalization
factors that arise when some fields propagate in different numbers
of dimensions (this is similar to the arguments used in the
context of neutrinos and other SM neutral states in the
case of large extra dimensions~\cite{AHDDMR}).

An illustrative example of the physics is provided by the following
slightly simplified model: Consider the 5th dimensional line segment
$y\in [0,\pi R/2]$ of the orbifold models with SU(5)-invariant
$O$-brane at $y=0$, and SU(5) breaking $O'$ brane at
$y=\ell\equiv \pi R/2$.  There are then three possible locations for
each multiplet: in the SU(5)-preserving 5d bulk, or at the $O$ or $O'$
branes.  If the light SM states arise from the zero modes $\Phi_0$
of a bulk field (generically denoted $\Phi$) then the normalization
factor of these zero modes is $1/\sqrt{M_*\ell}$,
where $M_*$ is the UV cutoff scale of our effective
higher-dimensional theory.\footnote{The
factor of $M_*$ arises from correctly normalizing the kinetic terms
so that the zero modes are fields with 4d canonical dimensions.}  
Alternatively some SM states can come from brane localized fields
-- generically $\psi$ -- which
we initially take to be located at $y=0$ for simplicity.  Then
the Yukawa interactions in the 4d effective Lagrangian are of the form
($x$ dependence suppressed):
\bea
{\cal L} &\simeq &\int dy\, \biggl\{ \la_0\de(y) \psi^3 + 
\la_1\de(y)\psi^2 \Phi(y) + \la_2\de(y)\psi \Phi^2(y) +
\la_3\de(y) \Phi^3(y) + \la_3' \Phi^3(y) \biggr\}\nonumber \\ 
&\simeq & \la_0 \psi^3 + 
{\la_1\over (M_*\ell)^{1/2}}\psi^2 \Phi_{(0)} + 
{\la_2\over (M_*\ell)}\psi \Phi_{(0)}^2 + 
{\la_3\over (M_*\ell)^{3/2}}\Phi_{(0)}^3 + {\la_1'\over (M_*\ell)^{1/2}}
\Phi_{(0)}^3 \, .
\label{eq:Yints}
\eea

This equation, which is crucial for the rest of our analysis,
requires some explanation.  For any interaction which contains at least
one brane localised field ($\la_0,\la_1$, or $\la_2$) there is
automatically a localising $\de(y)$-function when the Lagrangian
is written in 5d.  Then, for these terms, the number of factors
of $1/\sqrt{\ell}$ in the effective 4d Lagrangian is simply given
by the number of bulk fields in the interaction, as each one has this
wavefunction factor.  However for an interaction that involves only
bulk fields $\Phi$ there are two possibilities.  The first ($\la_3'$)
is that the Yukawa interaction occurs {\em everywhere} in the bulk.
In this case there is no $\de$-function, and for a cubic term two
of the wavefunction factors get cancelled by the integral over
$y\in [0,\ell]$, leading to just one power of $1/\sqrt{M_*\ell}$ overall.
The second possibility is that although the fields $\Phi$ live in the
bulk, the {\em interaction occurs only on the brane}.   This corresponds
to the $\de$-function in the $\la_3$ coupling, leading to three
factors of $1/\sqrt{M_*\ell}$ in the 4d effective coupling.  (In the
case with more than one brane such an interaction can occur at either,
or both of the branes.  This can be important in some circumstances.)

The second situation may seem unlikely, however it is in fact the typical
case. The reason for this is that, as we mentioned above, the minimal
supersymmetry in 5d corresponds to $N=2$ from the 4d perspective.  But
$N=2$ supersymmetry does not allow Yukawa interactions between matter
hypermultiplets, so the interactions between the component $N=1$ chiral
multiplets are confined to the branes where the $N=2$ SUSY is broken to
$N=1$ by the orbifold conditions. 

The various possible flavor models are characterized by the placement
of the $T_i$ and ${\bar F}_i$ multiplets that make up the three 
generations of quarks and leptons.   
In addition it is also possible for the Higgs multiplets to be located
in either the bulk or on the $O'$-brane --- if they were on the $O$-brane
the doublet-triplet splitting problem re-emerges. To have at least
some SU(5) mass relations, such as $b-\tau$ unification,  
the Higgs must reside in the SU(5)-symmetric bulk, in which
case they form components of ${\bf 5}$ and ${\bar {\bf 5}}$ in the bulk
which we denote $H$ and ${\bar H}$ respectively; for definiteness
we will focus on this case in the following.  In the next two
sections we explore forms for the Yukawa matrices following from
the preceding considerations of locality, geometry, and the spatially
dependent gauge and supersymmetries.  We will not impose extra
constraints, such as discrete or continuous flavor symmetries.

\section{Two-generation Models}
\label{sec:2gen}

For the charged fermion masses of the heaviest two generations there
are six dimensionless ratios: $m_\mu / m_\tau \simeq 1/17, m_s / m_b
\simeq 1/40, m_c / m_t \simeq 1/300,
V_{cb} \simeq 1/25, m_b / m_\tau \simeq 1, 
m_b /m_t \simeq 1/60$.  We aim to provide a qualitative understanding of the
first four of these in terms of the single volume factor $\ep \equiv 
1/ \sqrt{M_* \ell}$, according to $m_\mu / m_\tau, m_s/m_b, V_{cb}
\approx \ep$ and $m_c / m_t \approx \ep^2$.  In addition, we
begin with requiring that $m_b/m_\tau$ be precisely understood by
the SU(5) mass relation, and explore the most general forms of the
Yukawa matrices allowed by locality, without imposing extra constraints.
Later we will consider
the possibility of relaxing the requirement of precision $b-\tau$
unification.  In Section~5 we will also show the ratio
$m_b /m_t\simeq 1/60$ can be understood from geometry.

The up quark mass matrix in SU(5) is proportional to the coefficient
of an operator bilinear in the $T_i$ fields: $T_i T_j$.  A hierarchy of
eigenvalues requires that one $T$ is placed on a brane and the other
in the bulk.  The brane localized field has the heaviest up-type quark,
and is therefore to be identified with the third generation $T_3$.
To have a hope of
obtaining the SU(5) relation $m_b = m_\tau$ at the unification
scale, $T_3$ must be on $O$ rather than $O'$, so that the placement of
the two $T_i$ fields is unique.  The up quark Yukawa coupling matrix
for the two heavy generations, in the ``locality basis'', is
\beq
U\simeq\pmatrix{\de^3 & \de^2 \cr \de^2 & \ep \cr}.
\label{eq:ups}
\eeq
In this expression $\de$ corresponds to an SU(5)-violating parameter
of the same magnitude as the SU(5)-invariant parameter $\ep$.  We
emphasize that here and below we display only the hierarchical nature
of the Yukawa matrices, ignoring the order unity Yukawa couplings
of the 5d theory.  We
conclude that $m_c/m_t$ is necessarily of order $\ep^2$, and that
$V_{cb}$ necessarily contains a piece of order $\ep$ from
diagonalization of the up quark sector.

One interesting feature of Eq.(\ref{eq:ups}) is that because of the 
SU(5) violation $U$ is not symmetric as it would be in a 4d
SU(5) theory.  Also
note that the top quark Yukawa coupling has a magnitude proportional to
$\ep$.  This is not problematic, but shows that the
higher dimensional Yukawa coupling is closer to strong coupling than
in 4d theories.

The operators leading to down quark and charged lepton mass matrices
have the form $T_i \bar{F}_j$, leading to the 4d Yukawa matrices
$D_{ij}$ and $E_{ji}$.  These matrices will have a hierarchy of
order $\ep$ on the $i$ index due to the locality of the
$T_i$.  Since we require mass ratios $m_\mu / m_\tau, m_s/m_b \approx
\ep$, this implies that there is no additional hierarchy between
the two generations resulting from the index $j$:  $\bar{F}_2$ and
$\bar{F}_3$ must be located on branes of the same dimensionality.  Indeed
it is well-known that the large hierarchy in the up quark sector suggests
that, in SU(5) theories, the hierarchy is somehow associated with
the $T_i$ rather than with the $\bar{F}_j$.  In our scheme this difference
has a simple geometrical origin: the $T_i$ reside on branes with a
hierarchy of volumes, while the $ \bar{F}_j$ do not.

\begin{figure}
\begin{center}
\includegraphics[width=4in]{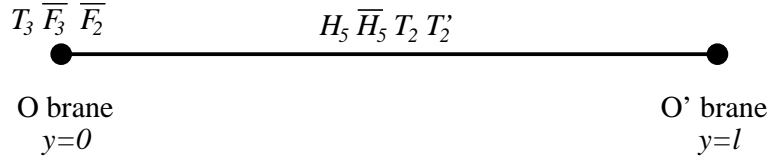}
\end{center}
\caption{The location of the Higgs and 2nd and 3rd generation
matter multiplets. The only alternative is for $\bar{F}_2$ to be
located at $O'$ instead of $O$.}
\label{fig:2gen}
\end{figure}

For locality to give the SU(5) $m_b/m_\tau$ mass relation, $\bar{F}_3$
must be located on the SU(5) brane $O$.  If it is in the bulk, then
$b$ and $\tau$ arise from different SU(5) multiplets  $\bar{F}_3$
and $ \bar{F}_3'$, so that they have unrelated Yukawa
couplings.  Hence we are forced to conclude that the $b$ and $t$ Yukawa
couplings are both of order $\ep$.  Thus in this 5d scheme the large $t/b$
mass ratio must arise from a large value for $\tan\beta$, the ratio of
electroweak VEVs, rather than from volume factors.  (At the end of
this section we discuss models without precision $b-\tau$ unification
where the large $m_t/m_b$ mass ratio is also explained by geometry.)
Thus our two generation theory is as shown in Figure~\ref{fig:2gen};
the only lack of uniqueness in the choice of location for the second
and third generations is whether $\bar{F}_2$ resides at $O$ or $O'$. 

With $ \bar{F}_{2,3}$ both on $O$, we may relabel the combination
which couples to $T_3$ as $ \bar{F}_3$, giving the Yukawa
couplings in the ``locality basis'':
\beq
D\simeq \pmatrix{\de^2 & \de^2 \cr 0 & \ep \cr} ,\qquad
E\simeq \pmatrix{\de^2 & 0 \cr \de^2 & \ep \cr} .
\label{eq:downs}
\eeq
The weak mixing $V_{cb}$ also acquires a contribution of order
$\ep$ from diagonalization of the down quark sector.  These order
$\ep$ contributions to quark flavor mixing, from both up and down
sectors, are simply a reflection of having $T_2$ in the bulk and $T_3$
on the brane.  Eqs.(\ref{eq:ups}) and (\ref{eq:downs}) together lead to
\beq
m_\mu/m_\tau \simeq m_s/m_b \simeq V_{cb} \simeq \ep,~~~
m_c/m_t \simeq \ep^2,~~{\rm and}~~m_b=m_\tau,
\label{eq:heavyrelns}
\eeq
as desired.  Indeed, it is remarkable, as can be seen from
Figure~\ref{fig:2gen}, that so much of the flavor structure of the
heaviest two generations is so simply understood by the single device
of putting $T_2$ in the bulk.  Finally note that there is no change
to these results even
if $\bar{F}_2$ is located at $O'$.  The operator  $T_3 \bar{F}_2
\bar{H}$ is now forbidden by locality, so that the Yukawa matrices
again take the form of Eq.(\ref{eq:downs}).

Before we discuss neutrino masses, let us mention an interesting
variation.  While the present configuration gives
$m_b=m_{\tau}$, it requires a large $\tan\beta$, as $m_b/m_t$ is
not volume-suppressed.  However, it is possible to get the right
$m_b/m_t$ without a large $\tan\beta$,  by putting both $\fb2$ and 
$\fb3$ in the bulk.  Since the locations of $T$'s are unaltered, 
the up-quark mass matrix Eq.(\ref{eq:ups}) stays the same, while the
down-quark and lepton matrices in Eq.({\ref{eq:downs}) become smaller
by a factor of $\ep$, which is good for $m_b/m_t$.  However, the
precise SU(5) relation $m_b=m_{\tau}$ is lost.  We will see how both
can be achieved in a 6d model.              

\subsection{Neutrino masses}

Neutrino masses can be generated via the see-saw mechanism if SU(5) 
singlet fields $N$ are introduced.  If the $N$'s are located at a 3-brane, 
they are expected to acquire Majorana mass of order $M_*$.  For bulk $N$'s,
contributions to Majorana masses from brane mass terms are of order $1/\ell$. 
In addition to this, we can write down bulk mass terms if we have more than
one hypermultiplet in the bulk, which gives masses of order $M_*$, 
but here, for simplicity, we assume that these large bulk masses are absent.

Now, as an interesting example, consider the case of a
single $N$ field. With $\bar{F}_2$ at $O$, large $\nu_\mu \nu_\tau$
mixing is inevitable: $V_{\mu \tau} \approx 1$.  The  $\bar{F}_{2,3}$
basis is defined by the couplings of $T_3$, so that the two couplings
$\bar{F}_{2,3} N H$ will be comparable independent of where
$N$ is located.  This is a striking
result: so much of the observed pattern of flavor in the heavy two
generations appear to emerge from putting all the matter on the brane
with SU(5) invariance, except for the field $T_2$.\footnote{
On the other hand, if we were to consider models with $\bar{F}_2$
and  $\bar{F}_3$ spatially
separated, both small and large mixing angle possibilities exist.  With
$N$ in the bulk $V_{\mu \tau} \approx 1$, while with $N$ on a brane
$V_{\mu \tau} \approx \ep^2$.  One's naive expectation that
quark-lepton unification will give $V_{\mu \tau} \approx V_{cb}
\approx \ep$ turns out not to be correct.  There is no
relation between the mixing of quark doublets (contained in $T$) and
lepton doublets (contained in $\bar{F}$) because SU(5) allows a 
different spatial arrangement for the $T$ and $\bar{F}$ fields.}

It would be exciting to obtain a neutrino mass enhanced from $v^2/M_G$ 
by an inverse power of $\ep$, corresponding to the measured atmospheric 
oscillation distance, but this does not seem to readily occur.  One
typically gets a suppression: $\ep^2 v^2/M_G$ for bulk $N$, or
$v^2/M_*$ for brane $N$.  Therefore it seems that
an additional ingredient is necessary to suppress the Majorana $NN$
mass.  For example, with $\bar{F}_3$ and $N$ both located on $O$, the
$NN$ mass might be suppressed if $(B-L)$ breaking occurs at $O'$ and is
mediated across the bulk by heavy fields.  If a second $N$ field is
placed in the bulk it would have a much larger Majorana mass, and
would give rise to a correspondingly smaller light neutrino
eigenvalue.  On extending the theory to three generations, this
eigenvalue can provide the smaller neutrino mass squared difference
required for solar neutrino oscillations.

\section{Three-generation Models in Six Dimensions}

Attractive three generation flavor models where all large
hierarchies are explained by geometry can be constructed using
a simple extension of the $S^1/(Z_2\times Z'_2)$ theories to
6 dimensions.  Specifically consider a 2d ``rectangular''
extra dimensional space with coordinates $y_1,y_2$ formed by the
product of two $S^1/(Z_2\times Z'_2)$ structures at right angles.
Equivalently this is just a 2d torus $T^2$ modded out by
$(Z_2\times Z'_2)^2$.  This rectangle has 4 orbifold 4-branes at
its border, $O_1$ along the $y_1$ axis, $O_2$ along the $y_2$ axis,
$O_1'$ along $y_2 = \ell_2$, and $O_2'$ along $y_1 = \ell_1$.
We take $\ell_2>\ell_1$.

This 6d system has a richer set of
possible symmetry structures than the 5d case case.  In particular
the two sets of orbifold actions, $(Z_2\times Z'_2)_1$ acting in the $y_1$
direction and $(Z_2\times Z'_2)_2$ acting in the $y_2$ direction, can
have different actions in field space:
\begin{itemize}
\item[A:]
The most immediate
extension of the 5d models is if for each factor, $Z_2$ commutes
with SU(5) and $Z_2'$ commutes with only SU(3)$\times$SU(2)$\times$U(1).
In this case the $O_1$ and $O_2$ branes preserve SU(5) with
(from a 4d counting) $N=2$ SUSY, while the $O_1'$ and $O_2'$ branes
preserve just SU(3)$\times$SU(2)$\times$U(1).  This symmetry
structure is illustrated in Figure~\ref{fig:6dgauge}A.
\item[B:]
Other options arise if we reduce the amount of SU(5)
breaking on the orbifold.  If instead of equivalent actions in the
two directions we change the gauge properties of one of the factors
so that both $Z_2$ and $Z_2'$ commute with SU(5), we can have
a situation where each 4-brane is SU(5) preserving except $O_2'$, say,
which only preserves SU(3)$\times$SU(2)$\times$U(1).  This
is illustrated in Figure~\ref{fig:6dgauge}B.
\end{itemize}
\begin{figure}
\begin{center}
\includegraphics[width=4in]{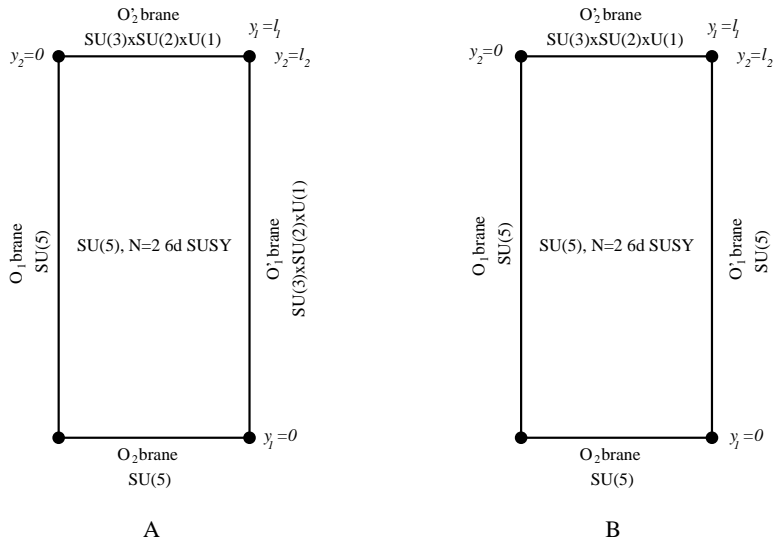}
\end{center}
\caption{Two of the possible structures of gauge and SUSY breaking
in 6d generalized from the $S^1/(Z_2\times Z_2')$ model in 5d.  In both
cases all 5d boundary branes possess $N=1$ 5d SUSY, while all 4d fixed-brane
corners have only 4d $N=1$ SUSY.}
\label{fig:6dgauge}
\end{figure}

Note that in both cases the bulk possesses the full SU(5) gauge symmetry,
but now with $N=2$ SUSY from the 6d perspective, equivalent to $N=4$ in
4d.  This ensures that the 6d bulk theory is free of gauge
anomalies~\cite{HMR2,HNOS}.
The corners of the rectangle are 3-branes which we will identify
with our 4-dimensional world.  The combination of orbifold
projections at these corners breaks the 6d $N=2$ SUSY all the way down
to $N=1$ 4d supersymmetry.

Another new feature of the 6d case is the possibility of placing
in different locations the fields giving rise to zero modes
that effectively fill out complete SU(5) multiplets.  For
example, in the $S^1/(Z_2\times Z_2')$ orbifold of Section~4
we needed to have both $T_2$ and $T_2'$ fields to realize zero
modes of matter in a full ${\bf 10}$ of SU(5), and moreover,
because there was only one extended 5d space (the full bulk in this
case) they necessarily had to be located together.  However in both 
6d cases illustrated in Figure~\ref{fig:6dgauge} there are now 4
fixed branes of spacetime dimensionality 5 on which we can place
the matter and explain hierarchies by volume factors. 
In some cases, such as the branes $O_1$ and $O_2$ of structure A,
or $O_1$ and $O_1'$ of structure B, the full SU(5) symmetry is realized
on both branes.  In such cases we are allowed to place, \eg, for structure
B,  $T_1$ on $O_1$, while $T_1'$ is located on $O_1'$.  As
we will discuss below this allows us to engineer some interesting
``texture zeros".
Alternatively we could utilize matter situated on the branes that
do not preserve SU(5), but only the SM subgroup.  In this
situation~\cite{HMR} it is only necessary that SM multiplets are
localized to the brane, and there is nothing that forces, say,
both the $\bar U$ and $\bar E$ components of what was formerly
combined in a $T$ to be placed on the same brane, or, as we will see, 
${\bar U}_1$ to be present at all if we wish to have a massless
up-quark. (It is also amusing to note that such constructions
in orbifold GUT theories allow exotic states with leptoquark
quantum numbers in a manner that is naturally consistent with the
stringent proton decay and flavor-changing constraints~\cite{exotic}.)

One worry concerning these 6d models is that the cutoff $M_*$ might
not be sufficiently far above the scales $1/\ell_{1,2}$ if the 6d gauge
theory quickly becomes strongly coupled.  If this were the case then
we would not be able to take the parameters $\ep_1=1/\sqrt{M_*\ell_1}$
and $\ep_2=1/\sqrt{M_*\ell_2}$ sufficiently small to be useful in
generating the observed flavor hierarchies.  However the usual naive
dimensional analysis or RG running estimate of the cutoff scale is
very dependent on the flat nature of the extra dimensions (with the usual
evenly spaced Kaluza-Klein spectrum).  As shown in Ref.~\cite{kaloper}
the form of the KK spectrum is highly dependent on the curvature
of the extra dimensional space (apart from the zero modes which
are unaffected since they arise for essentially topological reasons).
For example, replacing the $g=1$ torus $T^2$ with a simple 2d compact
hyperbolic manifold (CHM) (genus $g>1$ Riemann surfaces) leads to
an exponential squeezing of the excited KK modes to high scales.
(Specifically, the non-zero-mode KK states have masses that start at
$\exp(\al)/\ell$ rather than $m_{KK} \sim 1/\ell$, for some $\al$
which depends on the CHM chosen and which can easily be $O(10)$.)
On the other hand, it is simple to prove that the shape of the zero-mode
wavefunctions in the extra dimensions is still constant, and the
normalization factors for the zero-mode wavefunctions are still given
by $1/\sqrt{M_*\ell}$.  The raising of the excited KK spectrum
raises the scale at which the 6d gauge theory becomes strong, and
so allows larger values of $M_*\ell$. 

With these general considerations in mind we now turn to the
construction of a variety of three-generation flavor models
in 6d.  As well as the building of an attractive model our
interest is to demonstrate some of the possibilities of the
new framework.

\subsection{Two simple 3-generation models with small $\tan\be$}

One simple model based on the symmetry structure A of
Figure~\ref{fig:6dgauge} is depicted in Figure~\ref{fig:6dmodel}.
\begin{figure}
\begin{center}
\includegraphics[width=2.5in]{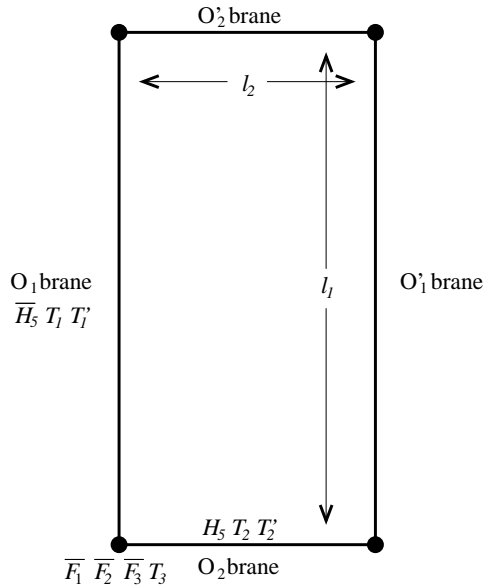}
\end{center}
\caption{The configuration of the MSSM Higgs and 3 generations
of matter multiplets in a simple 6D flavor model based on orbifold
structure A.}
\label{fig:6dmodel}
\end{figure}
As before, the quark and lepton mass hierarchies arise
from the $T$'s: all three generations $\fb1,\fb2,\fb3$ of matter
$\bar{\bf 5}$'s are located at the 4D intersection of $O_1$ and $O_2$, 
together with the third generation $\t3$, ensuring bottom-tau
unification.  On the other hand, $\t1+\t1'$ and $\t2+\t2'$ are located
on the 5D spaces $O_1$ and $O_2$, respectively.  Finally, $\bar{H}_5$
and $H_5$, containing the down-type and up-type Higgs multiplets, also
propagate on $O_1$ and $O_2$ respectively.  This configuration gives
rise to the following textures for Yukawa couplings in the
locality basis:
\beq
U\simeq \epsilon_2 
\pmatrix{\de_1^2 &\de_1 \de_2  & \de_1 \cr 
  \de_1 \de_2 &\de_2^2 &\de_2 \cr  
  \de_1  & \de_2  & 1},~~
D\simeq 
\epsilon_1\pmatrix{\de_1 & \de_1 & \de_1 \cr 
0 & \de_2 & \de_2 \cr 
0 &0 &1},~~ 
E  \simeq  
\epsilon_1\pmatrix{\de_1 & 0 & 0 \cr 
\de_1 & \de_2 &0 \cr
 \de_1 & \de_2 & 1}.
\label{eq:3matrices}
\eeq
Here $(\de_1,\ep_1)$ and $(\de_2,\ep_2)$ are volume suppressions
of order $1/\sqrt{M_*\ell_1}$ and $1/\sqrt{M_*\ell_2}$ associated
with the spaces $O_1$ and $O_2$,
respectively.  As before, the $\ep$'s and the $1$'s in
Eq.~(\ref{eq:3matrices}) are SU(5) preserving while the $\de$'s are
SU(5) violating.  Also, we have relabelled
$\fb1$, $\fb2$ and $\fb3$ such that
$T_3$ couples only to $\fb3$ while $T_2$ couples to $\fb2$
and $\fb3$.  Now, by taking $\ell_1 \gg \ell_2$, we obtain a
mass hierarchy for the light two generations, with the hierarchy again
strongest for the up-type quarks.  Altogether we have
$m_t:m_c:m_u \sim 1:\de_2^2:\de_1^2$ and
$m_{b,\tau}:m_{s,\mu}:m_{d,e}\sim 1:\de_2:\de_1$.  The placement
of the Higgs multiplets also introduces the hierarchy
$\lambda_b/\lambda_t \sim \de_1/\de_2$, so in this model 
$m_b/m_t$ is volume suppressed and large $\tan\be$ is not necessary.
Also note that we do not have the phenomenologically
disfavored SU(5) mass relations like
$m_d/m_s = m_e/m_u$, because the light generations propagate on
4-branes and are sensitive to the orbifold breaking of SU(5).
Thus even this quite simple model has many attractive features.

\begin{figure}
\begin{center}
\includegraphics[width=2.5in]{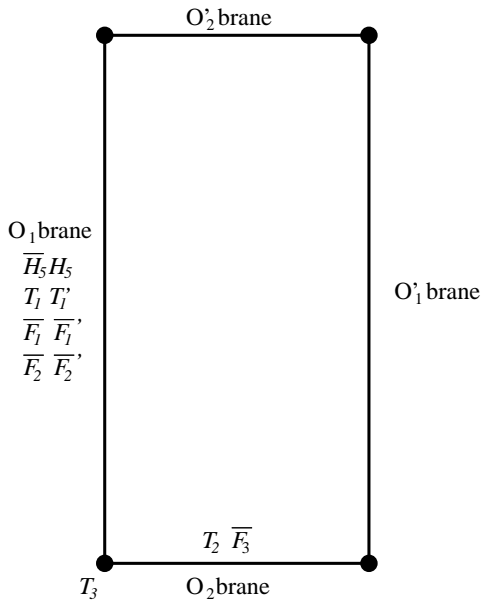}
\end{center}
\caption{The configuration of the MSSM Higgs and 3 generations
of matter multiplets in a simple 6d flavor model based on orbifold
structure B.}
\label{fig:var3alt}
\end{figure}

Much of the discussion on neutrino masses from
Section~\ref{sec:2gen} also carries over here.  Because the
$\overline{F}$'s share the same
location, large mixing angles are expected regardless of where the
$N$'s propagate.  (The small value of $\theta_{13}<.15$
\cite{Apollonio:1999ae} should in this model arise as
a mild accidental cancellation.)  If each $N$
propagates in the same space, this setup provides a simple realization
of the neutrino mass anarchy scenario \cite{Hall:2000sn}.
As before, the Majorana mass for an $N$ localized to the 3-brane may be
suppressed by distant breaking of $(B-L)$, and if a different $N$
propagating in the bulk or on a 4-brane has direct contact with the
breaking, it will have a larger Majorana mass, leading to a smaller
mass eigenvalue that could be relevant for solar neutrino oscillations. 

To illustrate how the gauge symmetry structure of the orbifold can impact
flavor model building, consider another simple model, this time based
on structure B (see Figure~\ref{fig:var3alt}).
Since the $O_2$ brane now feels no SU(5) breaking from orbifold
projections at either of its endpoints, it is no longer necessary
to include the fields $T_2'$ and $\fb3'$ so as to give the required
SM state.  As a consequence this construction gives another method of
realizing bottom-tau unification even though $\fb3$ propagates on a
4-brane, as there is just one $\bar{\bf 5}$ state, and thus one Yukawa
coupling for both $b$ and $\tau$.
Unwanted SU(5) mass relations are still avoided for the light two
generations because $\fb1$ and $\fb2$ contact the 4-brane $O_2'$ that only
preserves SU(3)$\times$SU(2)$\times$U(1).  By the same token, the
Higgs must be located on the vertical line to get doublet-triplet
splitting.  As before, $m_b/m_t$ is suppressed by a volume factor.
The resulting mass matrices, after appropriately relabelling $\overline{F}_1$
 and $\overline{F}_2$, are
\beq
U\simeq \epsilon_1 
\pmatrix{\de_1^2 &\de_1 \ep_2  & \de_1 \cr 
  \de_1 \ep_2 &\ep_2^2 &\ep_2 \cr  
  \de_1  & \ep_2  & 1},~~~ 
D\simeq E^T \simeq
\epsilon_1\pmatrix{\de_1^2 & \de_1^2 & \de_1 \ep_2 \cr 
0 & \de_1 \ep_2 & \ep_2^2 \cr 
\de_1 &\de_1 & \ep_2}.
\label{eq:var3}
\eeq 
Thus, this model is as attractive as the previous one.

These two models are not perfect, however.  They lead to
$\theta_c \sim m_d/m_s \sim m_e/m_u
\sim \de_2/\de_1$, when the correct numerical values are $\sim
1/5$, $1/20$, and $1/200$, respectively.  Of course because  
there are unknown coefficients contained in each element of the Yukawa
matrices it is not impossible that these ratios are corrected, but this
deviation from our philosophy is unappealing.  Nevertheless,
the textures of Eqs.(\ref{eq:3matrices}) and (\ref{eq:var3})
do go a long way towards explaining hierarchies of fermion masses
and mixings.  We now show it is possible to build on this basic
idea to do better.  

\subsection{Improved 3-generation models}

Consider a variation of the previous A-type model, shown in
Figure~\ref{fig:var2}.  This demonstrates how texture zeros can arise
in our framework (without giving a massless fermion), and also shows
how one can improve on some of the mass and mixing-angle relations
obtained from Eq.~(\ref{eq:3matrices}) in the original case.
\begin{figure}
\begin{center}
\includegraphics[width=2.5in]{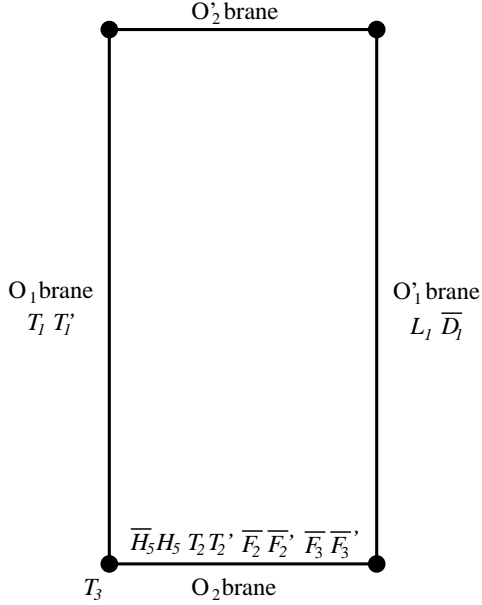}
\end{center}
\caption{A configuration based on orbifold structure A that gives
$D_{11}=E_{11}=0$, leading to $\theta_c\sim\sqrt{m_d/m_s}$.}
\label{fig:var2}
\end{figure}
The idea of this model follows from the well-known fact that
for the light generations of down-type quarks, the texture
\beq
D\simeq \pmatrix{0 & \de \cr \de & 1} ,\qquad
\label{eq:zero11}
\eeq
leads to the successful relation
$\theta_c \sim \sqrt{m_d/m_s}$ rather than $\theta_c \sim
m_d/m_s$ (provided that larger mixing does not come from the up
sector).  Moreover, if the $E$ Yukawa texture
has the same form, except that its (2,2) entry happens to
be somewhat larger, then the texture zero leads to a suppression of
the ratio $m_e/m_\mu$ relative to $m_d/m_s$ by $(D_{22}/E_{22})^2$. 
So if the above form of the Yukawa matrix is generated for $D$ and $E$,
the appropriate ratios of $\theta_c$, $m_d/m_s$, and $m_e/m_u$ can arise
merely from a factor of three difference between $D_{22}$ and
$E_{22}$~\cite{GJ} which is acceptable in our framework.

The configuration of Figure~\ref{fig:var2} realizes this texture by
spatially separating $T_1$ and $\fb1$, giving the Yukawa matrices 
\beq
U\simeq \epsilon_2 
\pmatrix{\de_1^2 &\de_1 \de_2  & \de_1 \cr 
  \de_1 \de_2 &\de_2^2 &\de_2 \cr  
  \de_1  & \de_2  & 1},~~~ 
D\simeq E^T \simeq
\epsilon_2\pmatrix{0 & \de_1 \delta_2 & \de_1 \delta_2 \cr 
\delta_2 \delta_1 & \de_2^2 & \de_2^2 \cr 
0 &0 & \delta_2},
\label{eq:var2}
\eeq
in the locality basis.  Note that the $(1,1)$ and $(3,1)$ entries
of $D$ and $E^T$ vanish because $\t1$ and $\t3$ are localized
away from $L_1$ and ${\bar D}_1$.  On the other hand the zero
in the $(3,2)$ entry of $D$ and $E^T$ is just due to the freedom to
relabel the combination of $\fb2$ and $\fb3$ that couples to $\t3$.
Thus this model predicts the desirable relation $\th_c\sim\sqrt{m_s/m_s}$.
Another nice feature of this model is that $m_b/m_t$ is suppressed by a
volume factor since $\fb3'$ lives on the $O_2$ 4-brane.  However, precision
bottom-tau unification is an accident and the model also
predicts $m_u/m_c \sim m_d/m_s$ when in fact $m_d/m_s$ is at
least a factor $\sim 10$ larger.

This last difficulty is avoided in an interesting way by
the model of Figure~\ref{fig:var1}.  
\begin{figure}
\begin{center}
\includegraphics[width=2.5in]{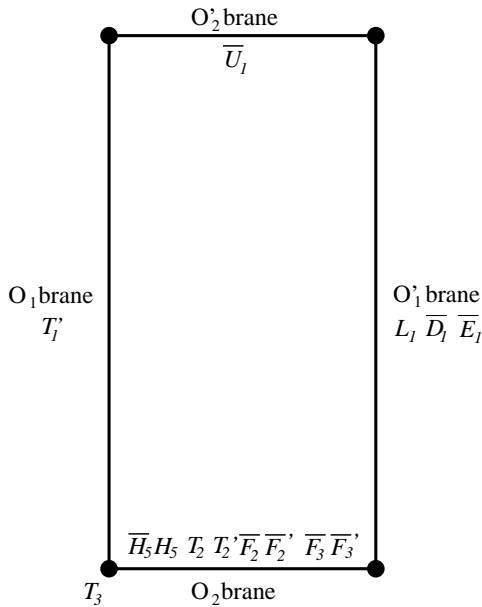}
\end{center}
\caption{A configuration based on orbifold structure A
that gives $\th_c\sim\sqrt{m_d/m_s}$ and $m_u=0$.}
\label{fig:var1}
\end{figure}
We have removed $T_1$ and distributed $\overline{U}_1$ and $\overline{E}_1$
onto $O_2'$ and $O_1'$ respectively.  Recall form Tables~1 and 2 that
each $T$ on an SU(5) preserving 4-brane (with at least one
SU(5)-violating boundary brane) contains massless zero modes
only for $\overline{U}$ and
$\overline{E}$, while each such $T'$ contains a massless zero mode only for
$Q$, and recall also that $O_1'$ and $O_2'$ preserve just
SU(3)$\times$SU(2)$\times$U(1).  The associated Yukawa matrices are thus
\beq
U\simeq \ep_2 
\pmatrix{0 &\de_1 \de_2  & \de_1 \cr 
  0 &\de_2^2 & \de_2 \cr  
  0  & \de_2  & 1},~~~
D\simeq E^T \simeq\ep_2
\pmatrix{0 & \de_1\de_2 & \de_1\de_2 \cr 
\de_1\de_2 & \de_2^2 & \de_2^2 \cr 
0 & 0 & \de_2},
\label{eq:var5}
\eeq
in the locality basis. 
The most striking and interesting
feature of this model is that it realizes in a simple way two features,
$m_u=0$ and $\th_c\sim\sqrt{m_d/m_s}$.  This is especially so
because $m_u=0$ is a solution to the strong CP problem (as
$\bar{U}_1$ has no Yukawa interactions), which may be consistent with
chiral perturbation theory~\cite{Kaplan:1986ru}.  Moreover, the hierarchy
$m_t/m_b$ is still explained by volume suppression in this model,
again because $\fb3'$ propagates on a 4-brane.  Therefore, except for
the fact that precision bottom-tau unification must still be regarded
as a numerical accident, this is a notably successful model.

%(An alternate arrangement has
%$\fb3$ on the 3-brane with $T_3$, and the first two generation
%$\overline{F}$'s on the $O_2$ brane. This preserves bottom-tau
%unification but gives $m_s/m_b \sim \de_2^2$ rather than
%$\de_2$.)

\section{Conclusions}

We have argued that 5 and 6-dimensional supersymmetric orbifold
GUT models allow an appealing explanation of the observed
hierarchical structure of the quark and lepton masses
and mixing angles.  Our hypothesis is that flavor hierarchies
arise from the geometrical suppression of some couplings due to
wavefunction normalization factors when fields propagate in
different numbers of dimensions.  Moreover, if fields propagate
on different fixed branes, restrictions arising from locality in
the extra dimension allow interesting texture zeroes to be simply
explained.  In addition the detailed nature of the SU(5)-breaking
orbifold projections lead to simple theories where $b-\tau$ unification
is maintained but similar, disfavored, SU(5) relations for the lighter
generations are naturally avoided.  We find that simple 5d models based
on $S^1/(Z_2\times Z_2')$ are strikingly successful in explaining many
features of the masses and mixing angles of the 2nd and 3rd generation,
this success resulting from the single simple assumption that    
the only matter propagating in the 5th dimension is the second
generation ${\bf 10}$ of SU(5).  These ideas were then extended 
in Section~5 to successful three generation models of flavor,
constructed by generalizing the $S^1/(Z_2\times Z'_2)$ model to
six dimensions on orbifolds with structure $T^2/(Z_2\times Z'_2)^2$.
Once again the primary hypothesis leading to attractive models was the
distribution of the three generations of ${\bf 10}$'s, $\t1,\t2,\t3$
on branes of different co-dimensionality or linear extent in the
extra dimensions.  
Some novel features of these models, including a simple $m_u=0$
configuration leading to a solution of the strong CP problem were
also discussed.

Finally, a valid criticism of our models is that they only give the
hierarchical structure of the quark and lepton masses and mixing
angles and do not allow precise predictions because of the many
unknown $O(1)$ Yukawa 
couplings of the high-scale theory.  In this regard our models share
features of the Froggert-Nielsen mechanism~\cite{FN}, where such
uncertainties are also present.
We believe that by utilizing further features unique to
higher-dimensional GUT theories it is possible to fix many
of these $O(1)$ parameters, thus leading to a set of precise predictions
for relations among the low-energy quark and lepton masses and mixing
angles~\cite{future}.  Certainly many interesting issues raised
by the success of the $S^1/(Z_2\times Z_2')$ orbifold GUT models and
its generalizations remain to be investigated.

\section*{Acknowledgements}
We are grateful to Arthur Hebecker for discussions.  JMR thanks the 
Theory Group, LBNL, for their kind hospitality during this work.


\begin{thebibliography}{99}

\bibitem{GG}
H. Georgi and S. Glashow, Phys. Rev. Lett. {\bf 32} (1974) 438;\\
H. Georgi, Unified Gauge Theories
in {\em Proceedings, Coral Gables 1975, Theories and Experiments
In High Energy Physics}, New York (1975).

\bibitem{PS}
J.~Pati and A.~Salam, Phys. Rev. {\bf D8} (1973) 1240;
{\em ibid} {\bf D10} (1974) 275.

\bibitem{GQW}
H. Georgi, H. Quinn and S. Weinberg, Phys. Rev. Lett. {\bf 33} (1974) 451.

\bibitem{DRW}
S. Dimopoulos, S. Raby and F. Wilczek, Phys. Rev. {\bf D24} (1981) 1681.

\bibitem{DG}
S. Dimopoulos and H. Georgi, Nucl. Phys. {\bf B193} (1981) 150.

\bibitem{unif}
N.~Sakai, Z. Phys. {\bf C11} (1981) 153;\\
L.~Ibanez and G.~Ross, Phys. Lett. {\bf B105} (1981) 439;
{\em ibid} {\bf B110} (1982) 215;\\
M.~Einhorn and D.R.T.~Jones, Nucl. Phys. {\bf B196} (1982) 475;\\
W.~Marciano and G.~Senjanovic, Phys. Rev. {\bf D25} (1982) 3092;\\
P.~Langacker, M.-X.~Luo, Phys. Rev. {\bf D44} (1991) 817;\\
J.~Ellis, S.~Kelley and D.~Nanopoulos, Phys. Lett. {\bf B260} (1991) 131;\\
U.~Amaldi, W.~de~Boer and H.~Furstenau, Phys. Lett. {\bf B260} (1991) 447.

\bibitem{BEGN}
M.~S.~Chanowitz, J.~Ellis and M.~K.~Gaillard, Nucl.\ Phys.\ B {\bf 128}
(1977) 506;\\
A.~J.~Buras, \etal,
Nucl.\ Phys.\ B {\bf 135} (1978) 66.

\bibitem{GJ}
H.~Georgi and C.~Jarlskog,
Phys.\ Lett.\ B {\bf 86} (1979) 297.
%%CITATION = PHLTA,B86,297;%%

\bibitem{FN}
C.~D.~Froggatt and H.~B.~Nielsen,
Nucl.\ Phys.\ B {\bf 147} (1979) 277.
%%CITATION = NUPHA,B147,277;%%

\bibitem{textures}
J.~Harvey, D.~Reiss and P.~Ramond,
Nucl.\ Phys.\ B {\bf 199} (1982) 223;\\
%%CITATION = NUPHA,B199,223;%%
S.~Dimopoulos, L.~Hall and S.~Raby,
Phys.\ Rev.\ Lett.\  {\bf 68} (1992) 1984;
%%CITATION = PRLTA,68,1984;%% 
Phys.\ Rev.\ D {\bf 45} (1992) 4192;\\
%%CITATION = PHRVA,D45,4192;%%
H.~Arason, \etal,
Phys.\ Rev.\ D {\bf 47} (1993) 232, [hep-ph/9204225];\\
%%CITATION = HEP-PH 9204225;%%
P.~Ramond, R.~Roberts and G.~Ross,
Nucl.\ Phys.\ B {\bf 406} (1993) 19, [hep-ph/9303320];\\
%%CITATION = HEP-PH 9303320;%%
G.~Giudice,
Mod.\ Phys.\ Lett.\ A {\bf 7} (1992) 2429, [hep-ph/9204215];\\
%%CITATION = HEP-PH 9204215;%%
L.~Ibanez and G.~Ross,
Phys.\ Lett.\ B {\bf 332} (1994) 100, [hep-ph/9403338].
%%CITATION = HEP-PH 9403338;%%

\bibitem{BB}
See, \eg, K.~Babu and S.~Barr,
Phys.\ Rev.\ D {\bf 50} (1994) 3529,
[hep-ph/9402291].
%%CITATION = HEP-PH 9402291;%%

\bibitem{MP}
For a recent treatment: H.~Murayama and A.~Pierce,
[hep-ph/0108104].
%%CITATION = HEP-PH 0108104;%%

\bibitem{kaw}
Y.~Kawamura, Prog. Theor. Phys. {\bf 103} (2000) 613,
[hep-ph/9902423]; [hep-ph/0012125]; [hep-ph/0012352].

\bibitem{AF}
G.~Altarelli and F.~Feruglio,
Phys.\ Lett.\ B {\bf 511} (2001) 257,
[hep-ph/0102301].
%%CITATION = HEP-PH 0102301;%%

\bibitem{HN}
L.~Hall and Y.~Nomura,
Phys.\ Rev.\ D {\bf 64} (2001) 055003,
[hep-ph/0103125].
%%CITATION = HEP-PH 0103125;%%

\bibitem{HMR}
A.~Hebecker and J.~March-Russell, [hep-ph/0106166].
%%CITATION = HEP-PH 0106166;%%

\bibitem{HMR2}
A.~Hebecker and J.~March-Russell, [hep-ph/0107039].
%%CITATION = HEP-PH 0107039;%%

\bibitem{recent}
R.~Barbieri, L.~Hall and Y.~Nomura, [hep-th/0107004];\\
%%CITATION = HEP-TH 0107004;%%
J.~Bagger, F.~Feruglio and F.~Zwirner,
[hep-th/0107128];\\
%%CITATION = HEP-TH 0107128;%%
A.~Masiero, \etal,
[hep-ph/0107201];\\
%%CITATION = HEP-PH 0107201;%%
C.~Csaki, G.~Kribs and J.~Terning,
[hep-ph/0107266];\\
H.~Cheng, K.~Matchev and J.~Wang,
[hep-ph/0107268];\\
%%CITATION = HEP-PH 0107268;%%
L.~Hall, H.~Murayama and Y.~Nomura,
[hep-th/0107245];\\
%%CITATION = HEP-TH 0107245;%%
N.~Haba, \etal, [hep-ph/0108003];\\
%%CITATION = HEP-PH 0108003;%%
T.~Asaka, W.~Buchmuller and L.~Covi,
[hep-ph/0108021];\\
%%CITATION = HEP-PH 0108021;%%
N.~Borghini, Y.~Gouverneur and M.~Tytgat,
[hep-ph/0108094];\\
%%CITATION = HEP-PH 0108094;%%
T.~Li, [hep-th/0107136];
%%CITATION = HEP-TH 0107136;%%
[hep-ph/0108120];\\
%%CITATION = HEP-PH 0108120;%%
R.~Dermisek and A.~Mafi, [hep-ph/0108139].
%%CITATION = HEP-PH 0108139;%%

\bibitem{HNOS}
L.~Hall, Y.~Nomura, T.~Okui and D.~Smith,
[hep-ph/0108071].
%%CITATION = HEP-PH 0108071;%%

\bibitem{string}
P. Candelas, \etal, Nucl. Phys. {\bf B258} (1985) 46;\\
E.~Witten, Nucl. Phys. {\bf B258} (1985) 75.\\
For a recent discussion see: A.~Faraggi, [hep-ph/0107094].
%%CITATION = HEP-PH 0107094;%%

\bibitem{CPRT}
R.~Contino, L.~Pilo, R.~Rattazzi and E.~Trincherini,
[hep-ph/0108102].
%%CITATION = HEP-PH 0108102;%%

\bibitem{Kaplan:1986ru}
D.~Kaplan and A.~Manohar,
Phys.\ Rev.\ Lett.\  {\bf 56}, 2004 (1986).
%%CITATION = PRLTA,56,2004;%%

\bibitem{Apollonio:1999ae}
M.~Apollonio {\it et al.}  [CHOOZ Collaboration],
Phys.\ Lett.\ B {\bf 466}, 415 (1999)
[hep-ex/9907037].
%%CITATION = HEP-EX 9907037;%%

\bibitem{Hall:2000sn}
L.~Hall, H.~Murayama and N.~Weiner,
Phys.\ Rev.\ Lett.\  {\bf 84}, 2572 (2000)
[hep-ph/9911341].
%%CITATION = HEP-PH 9911341;%%

\bibitem{AHDDMR}
N.~Arkani-Hamed, \etal, [hep-ph/9811448];
%%CITATION = HEP-PH 9811448;%% 
[hep-ph/9903224];\\
%%CITATION = HEP-PH 9903224;%%
K.~R.~Dienes, E.~Dudas and T.~Gherghetta,
Nucl.\ Phys.\ B {\bf 557} (1999) 25, [hep-ph/9811428].
%%CITATION = HEP-PH 9811428;%%

\bibitem{exotic}
K.~Babu, \etal,
Phys.\ Lett.\ B {\bf 402} (1997) 367,
[hep-ph/9703299];\\
%%CITATION = HEP-PH 9703299;%%
G.~Altarelli, \etal,
Nucl.\ Phys.\ B {\bf 506} (1997) 3, [hep-ph/9703276].
%%CITATION = HEP-PH 9703276;%%

\bibitem{kaloper}
N.~Kaloper, \etal, Phys.\ Rev.\ Lett.\  {\bf 85} (2000) 928,
[hep-ph/0002001].
%%CITATION = HEP-PH 0002001;%%

\bibitem{future}
Work in progress.

\end{thebibliography}
\end{document}